\documentclass[sigconf,nonacm]{acmart}

\usepackage{booktabs} 

\begin{document}
\title{Physically-based Path Tracer using WebGPU and OpenPBR}

\author{Simon Stucki}
\orcid{0009-0006-1172-9165}
\affiliation{%
  \institution{ZHAW Zurich University of Applied Sciences}
  \city{Winterthur}
  \country{Switzerland}}
\email{stucksi1@students.zhaw.ch}

\author{Philipp Ackermann}
\orcid{0000-0003-4607-7620}
\affiliation{%
  \institution{ZHAW Zurich University of Applied Sciences}
  \city{Winterthur}
  \country{Switzerland}}
\email{philipp.ackermann@zhaw.ch}

\begin{abstract}
This work presents a web-based, open-source path tracer for rendering physically-based 3D scenes using WebGPU and the OpenPBR surface shading model. While rasterization has been the dominant real-time rendering technique on the web since WebGL's introduction in 2011, it struggles with global illumination. This necessitates more complex techniques, often relying on pregenerated artifacts to attain the desired level of visual fidelity. Path tracing inherently addresses these limitations but at the cost of increased rendering time. Our work focuses on industrial applications where highly customizable products are common and real-time performance is not critical. We leverage WebGPU to implement path tracing on the web, integrating the OpenPBR standard for physically-based material representation. The result is a near real-time path tracer capable of rendering high-fidelity 3D scenes directly in web browsers, eliminating the need for pregenerated assets. Our implementation demonstrates the potential of WebGPU for advanced rendering techniques and opens new possibilities for web-based 3D visualization in industrial applications.
\end{abstract}

%
%


%
%


\begin{teaserfigure}
\begin{center}
    \includegraphics[width=1.0\textwidth]{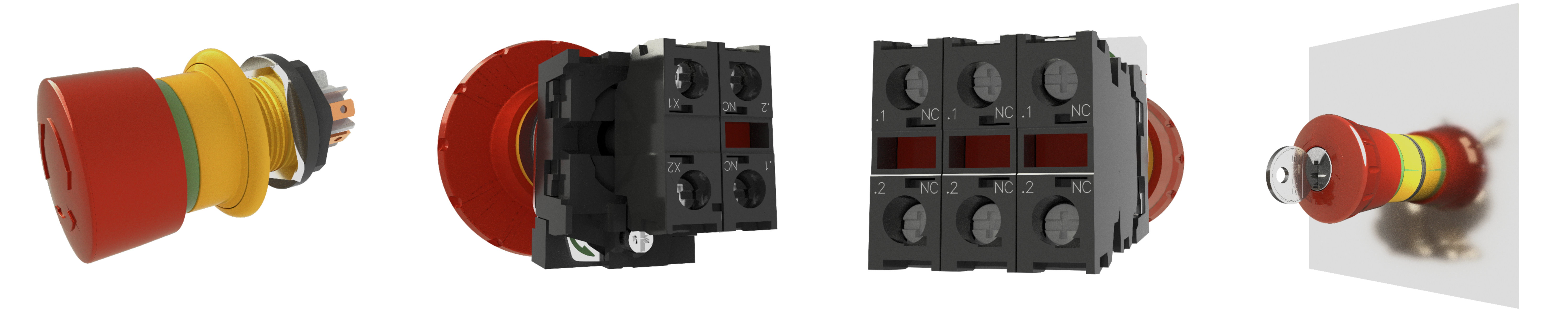}
\caption{Path traced renderings of pushbutton CAD models. From left to right: side view with metallic reflection, rear view exhibiting red color bleeding, rear view with ambient occlusion, and pushbutton mounted on a metallic surface, including the reflection of the Stanford bunny model \cite{turkLevoy1994} positioned in the scene.}
\Description{Four path traced renderings of CAD models showing metallic reflection, color bleeding, ambient occlusion, and shadows.}
\label{fig:teaser}
\end{center}
\end{teaserfigure}

\maketitle

\newpage
\section{Introduction}

Rasterization is a rendering technique that projects 3D scene geometry onto a 2D plane. Historically, the technique has been widely adopted in real-time rendering due to its efficiency. However, rasterization has limitations in achieving photorealism. One of the main limitations is the lack of support for global illumination. These phenomena can be observed in objects like mirrors or metallic surfaces. Effects such as refraction, reflection, and ambient occlusion require additional techniques.

Various methods have been developed to address these limitations, including pre-baked shadow, environment \cite{greene1986environment} and light maps; screen space reflections (SSR) \cite{screenSpaceReflectionsStackowiak}; screen space ambient occlusion (SSAO) \cite{bavoil2008ssao}; and screen space directional occlusion (SSDO) \cite{ritschel2009ssdo}.

These approaches induce complexity, may need to be computed at the assembly level, and can be computationally expensive. An alternative rendering technique resembles reality more closely and inherently alleviates these limitations: ray tracing.

\subsection{Ray Tracing}

Ray tracing is a powerful technique for rendering photorealistic scenes including global illumination. Historically, ray tracing has been mainly used in offline rendering due to its computational complexity. The theory behind ray tracing techniques is well established \cite{whitted2020OriginsOfGlobalIllumination}.

\newpage
Kajiya defined the rendering equation in 1986, coining the term path tracing. Generally, path tracing can be seen as an integration of the rendering equation, which is defined as:

\begin{displaymath}
  \label{eqn:rendering-equation}
  I(x, x') = g(x, x') [\epsilon(x, x') + \int_{S} p(x, x', x'')I(x', x'')dx'']
\end{displaymath}

\noindent where $I(x, x')$ is the radiance from point $x$ to point $x'$. $x$ for example being the camera position and $x'$ the intersected object. $g(x, x')$ is a geometry term determining how much light is transmitted. This depends on the distance and possibly occlusions such as transparent surfaces. $\epsilon(x, x')$ is the emitted radiance, generally used for light sources. The integral term is taken over $S$ = $\cup S_i$ which is the union of all surfaces. $p(x, x', x'')$ is the bidirectional reflection function, which describes how light from all possible directions $x''$ is reflected at the surface. \cite{kajiya1986rendering}

Ray tracing techniques focus on optimizing the rendering equation. Research in the 1990s focused on efficient light transport techniques such as bidirectional light transport and Metropolis light transport \cite{veachMonteCarloLightTransport}. Concurrently, production offline renderers like Blue Moon Rendering Tools (BMRT) were developed \cite{bmrt}.

Recent hardware advancements, including dedicated ray tracing acceleration \cite{evolutionOfGPU}, have made real-time ray tracing feasible, sparking increased research. Notable developments include reservoir-based spatio-temporal importance resampling (ReSTIR) \cite{restir} and subsequent improvements \cite{restirAdvancements,restirGeneralized}.

\section{Use Cases}

E-commerce represents a key use case for product renderings. Traditional catalogues struggle with highly configurable products. Computer graphics addresses this challenge through product configurators for virtual assembly and visualization.

Leveraging existing CAD models, prevalent in mechanical engineering and product design, for end-user applications offers a significant advantage. These models contain geometric and material information, which eliminates the need for redundant 3D models for marketing purposes. One example is EAO, which manufactures highly customizable industrial pushbuttons and operator panels as visualized in Figure~\ref{fig:teaser}. Due to the nature of the product, the number of possible assemblies grows almost exponentially with the number of components. To facilitate the configuration process, a web-based configurator is optimal because it can be used on a large variety of devices without having to install additional software. Given the company's priority for photorealistic renderings of the assemblies, using ray tracing techniques is a compelling choice.

Pre-rendering all product configurations is theoretically possible for a finite number of configurations, but computationally expensive. If this is not feasible, real-time rendering is an alternative. For real-time rendering, one option is remote rendering \cite{remoteRendering}, which employs a server to render the scene and stream the visualization to the browser. The main drawbacks of this approach are network latency, reliance on network stability, as well as operational cost for the server infrastructure, which frequently requires dedicated GPUs for rendering. Another option is client-side rendering. Most frequently, rasterization approaches are used for web-based renderings. However, due to the limitations in global illumination effects, the need for a real-time, client-side ray tracing solution for the web becomes apparent when considering the use case.

\begin{figure}[H]
  \includegraphics[width=0.9\columnwidth]{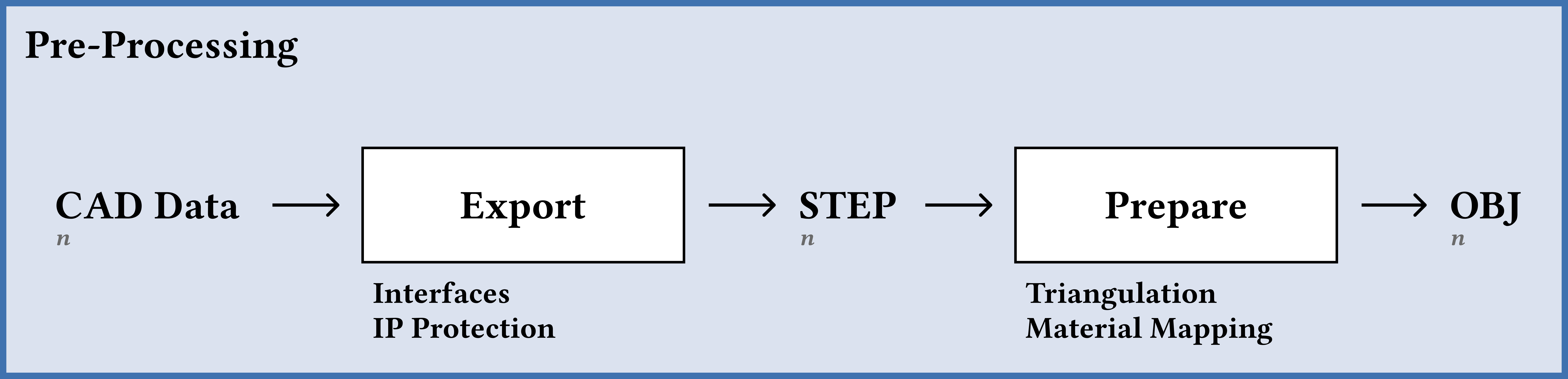}
  \caption{A two-step pre-processing stage is employed for offline as well as real-time rendering pipelines.}
  \Description{Diagram showing the steps in data pre-processing. The process begins with original CAD Data, which is then converted into a STEP file. This file is then further processed and converted into an OBJ file.}
  \label{fig:cad-preprocessing}
\end{figure}

CAD models require pre-processing as described in Figure~\ref{fig:cad-preprocessing}. This step involves surface triangulation, potential further refinement using algorithms used to generate level of detail (LOD) artifacts \cite{luebke2003level}, and measures to protect intellectual property (IP) rights by removing proprietary data. While numerous CAD formats include material information, it is often unsuitable for rendering. To address this, a material mapping can be defined, translating CAD materials to a suitable representation for the rendering pipeline.

\begin{figure}[H]
  \includegraphics[width=\columnwidth]{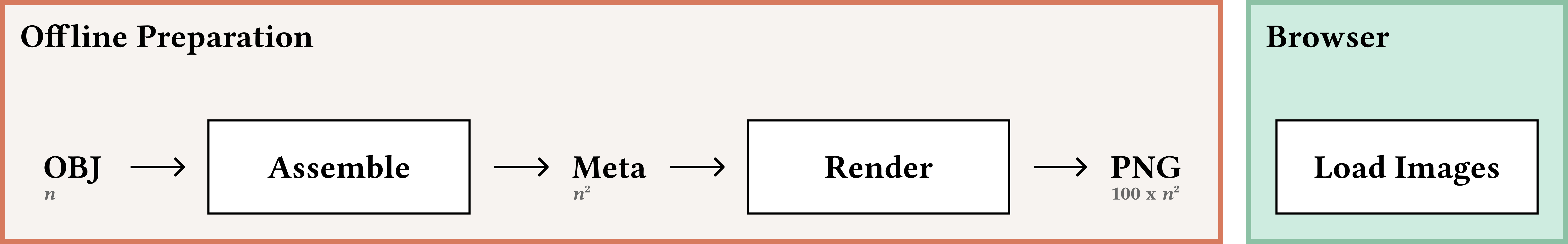}
  \caption{In an offline rendering setup, the number of images grows exponentially. The number of images increases further when offering 360° viewing.}
  \Description{Diagram showing offline rendering pipeline. It starts with OBJ files, which are then assembled into a meta object. This information is then used to generate images for each viewing angle.}
  \label{fig:cad-offline}
\end{figure}

An offline rendering pipeline generates static images of the assembly, which are then displayed in the browser. This means that all possible assemblies need to be rendered and stored upfront. As the number of components increases, the number of possible combinations grows exponentially, which can lead to large amounts of storage and processing power being required, as shown in Figure~\ref{fig:cad-offline}.

\begin{figure}[H]
  \includegraphics[width=\columnwidth]{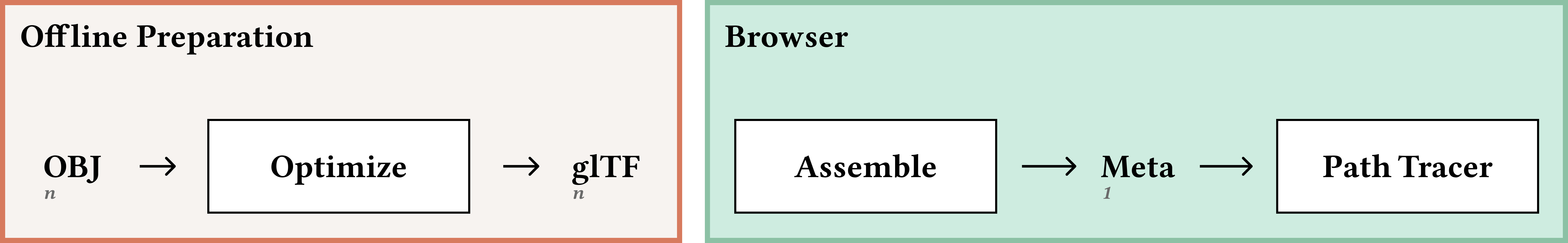}
  \caption{A real-time rendering pipeline solely relies on having an adequate model for each component of the assembly. It does not require pre-rendering every possible configuration.}
  \Description{Diagram showing real-time rendering pipeline. It starts with OBJ files, which are then optimized into glTF files for transport. The browser then uses these files to assemble meta information and render the scene using a path tracer.}
  \label{fig:cad-online}
\end{figure}

This is the main benefit of using a real-time rendering pipeline, as illustrated in Figure~\ref{fig:cad-online}. The rendering is done in the browser, which means that the server only needs to provide the geometry and material information in an exchange format such as glTF. This approach is more flexible and can be used for a larger number of configurations. The amount of data to be stored and processed offline grows linearly with the number of components, independent of the number of possible configurations.

\section{Prior Work}

There are a variety of path tracers available for the web, most of which are based on WebGL. The first experiments using WebGL for path tracing were implemented as early as 2010. One such example is the Google experiment demonstrating a Cornell Box \cite{goral1984modeling} with basic primitive shapes such as spheres and planes \cite{pathTracerWallace}. Since then, multiple open-source implementations for the web have been created. Some of the most widely known open-source web-based path tracers include:

\begin{itemize}
  \item {\texttt{three-gpu-pathtracer}} \cite{ThreeJsPathTracerJohnson}.
  \item{\texttt{Three.js PathTracer}} \cite{ThreeJsPathTracerLoftis}.
  \item {\texttt{dspbr-pt}} \cite{PathTracerDassault}.
\end{itemize}

WebGPU is a new web standard which, unlike WebGL, is no longer based on OpenGL. It is designed to be more efficient and flexible than its predecessor. Flexibility is mainly exhibited through its support for general-purpose GPU (GPGPU) computations by design.
Common 3D engines based on rasterization, including Babylon.js \cite{BabylonJSWebGPUSupport}, PlayCanvas \cite{playCanvasWebGPUSupport}, Three.js \cite{ThreeJSWebGPUSupport}, and Unity \cite{UnityWebGPUSupport}, have officially announced support for or started working on WebGPU.

Various applications of WebGPU have been investigated in recent years. Examples include Dynamical.JS, a framework for visualizing graphs \cite{dotson2022dynamicaljs}; RenderCore, a research-oriented rendering engine \cite{Bohak_Kovalskyi_Linev_Mrak_Tadel_Strban_Tadel_Yagil_2024}; and demonstrations of how to use WebGPU for client-side data aggregation \cite{kimmersdorfer2023webgpu}.

Research on the performance comparison between existing 3D engines and WebGPU engines has been conducted as part of the development of FusionRender, which concluded that measurable performance gains can be achieved by using WebGPU, but only when effectively leveraging its novel design principles \cite{fusionRenderWebGPU}. Similar findings have emerged from other independent studies, demonstrating that WebGPU can be faster than WebGL \cite{webGPUWebGis, fransson2023performance, CHICKERUR2024919}.

In light of these findings, WebGPU presents a transformative opportunity. It is particularly well-suited for the development of a new real-time path tracing library for the web.

\section{Physically Based Rendering}

Defining the geometry is one part of the equation. Another major part is defining materials. Materials determine how a surface interacts with light. The core idea of physically based rendering (PBR) is to simulate the interaction of light using physical models. Instead of fine-tuning parameters for a specific look and feel and having to adjust based on desired lighting situations, PBR aims to define a material that behaves consistently in different lighting situations. One prominent example of a PBR system is pbrt, an open-source renderer with an associated literate programming book \cite{Pharr_Physically_Based_Rendering_2023}.

Based on these principles, different material shading models have been developed. MaterialX is an open standard for representing materials using a node-based system. Originally developed by Lucasfilm in 2012, it has since been adopted by the Academy Software Foundation as an open standard. While not strictly limited to PBR, the standard provides a wide range of features for describing physically based materials \cite{Harrysson2019}.

OpenPBR, also hosted by the Academy Software Foundation as an open standard, differs from MaterialX by providing an uber shader approach rather than a node-based system. The uber shader approach is defined by a fixed set of inputs which can be adjusted but does not support custom node graphs like those in MaterialX. OpenPBR combines aspects of Autodesk Standard Surface and Adobe Standard Material. Its parameters allow for configuring metalness, glossy-diffuse, subsurface, coat, fuzz, emission, and more. To date, the standard has a reference implementation in MaterialX. \cite{openPBRSpec}

The emphasis on high-quality PBR materials is advantageous for the industrial use cases described in this paper. Furthermore, the goal of interoperability within the broader ecosystem strengthens the case for using OpenPBR in our implementation.

\section{Results}

The work presented in this paper introduces a web-based path tracer that utilizes WebGPU for near real-time rendering. The renderer supports loading glTF models and allows for the configuration of materials using OpenPBR parameters.

The renderer is available as an open-source project with extensive documentation at \url{https://www.github.com/StuckiSimon/strahl}. The library is published on the npm registry as \texttt{strahl}.

\subsection{Implementation}

\begin{figure}[H]
  \includegraphics[width=1.0\columnwidth]{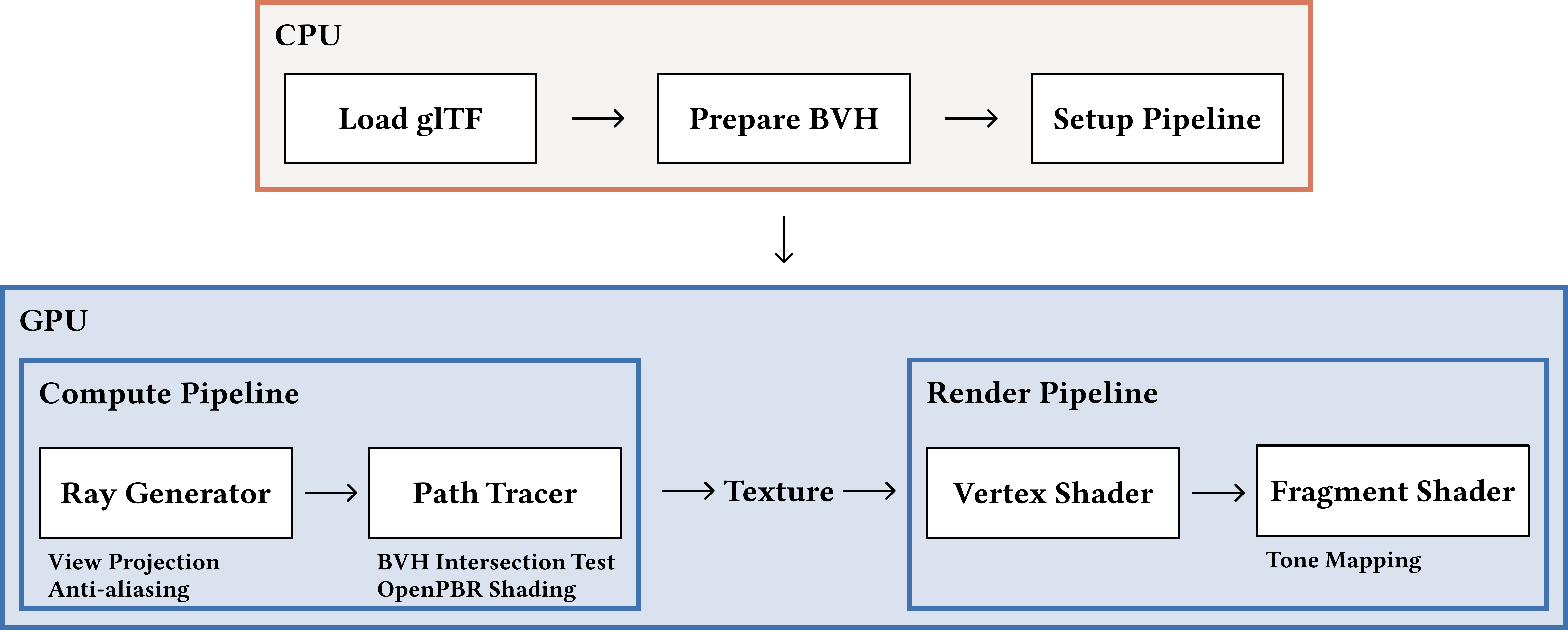}
  \caption{Path Tracer Pipeline, distinguishing GPU and CPU tasks. The stages of the compute pipeline and render pipeline are executed sequentially.}
  \Description{Diagram showing the path tracer pipeline. The pipeline starts with the CPU setup, which consists of loading the glTF model, preparing the BVH and setting up the pipeline. Then the GPU starts the compute pipeline by generating rays, tracing the ray and finally rendering the output in the render pipeline.}
  \label{fig:path-tracer}
\end{figure}

Figure~\ref{fig:path-tracer} illustrates the procedure of the path tracer. Scene preparation is performed on the CPU. This setup needs to be done once per visualization. Subsequent sampling of the scene is carried out repeatedly on the GPU, which constitutes the most computationally intensive part of the process.

\subsubsection{CPU Setup}

The CPU processing begins by loading the glTF model. As an acceleration structure for ray intersection tests, the renderer employs a bounding volume hierarchy (BVH) based on axis-aligned bounding boxes (AABB). The core concept is to group adjacent objects in a bounding volume and structure the hierarchy so that all child elements of a node are contained in the bounding volume of the parent node. This hierarchical structure allows for early rejection of branches not intersected by the ray, reducing the number of intersection tests required per ray cast from $O(n)$ to $O(log(n))$. Once the BVH is constructed, the scene is prepared for rendering and the pipeline is set up.

\subsubsection{Ray Generator}

The ray generator is responsible for casting rays into the scene according to the view projection. Based on the use case, the generator uses perspective projection. The path tracer employs a backward ray tracing approach, tracing rays from the camera into the scene. To mitigate aliasing effects, rays for the same pixel are slightly varied across multiple samples. Parallelization is achieved by computing adjacent pixels concurrently across multiple GPU cores.

To sample distributions, the path tracer selects random values using a random number generator (RNG). Several suitable pseudorandom number generators have been considered, including Mersenne Twister \cite{rngMersenneTwister} and Xorshifts such as the method described by Marsaglia \cite{marsaglia2003xorshift}. Our renderer uses a permuted congruential generator (PCG) as the RNG due to its good performance and quality \cite{o2014pcg}.

\subsubsection{Path Tracer}

Afterwards, the core component of the path tracer recursively traces the rays. The procedure checks for intersections using the BVH and calculates the radiance contribution of the intersected objects based on the OpenPBR material parameters. The OpenPBR standard employs an uber shader approach, which provides a balance between flexibility and performance. The OpenPBR support is focused on the most relevant features for industrial use cases, permitting realistic rendering of metallic, glossy, and diffuse surfaces.

Importance sampling is used to reduce the number of samples required for accurate results by prioritizing sample directions that contribute more significantly to the final image. For highly metallic surfaces, sampling is steered towards the reflection direction, while for diffuse surfaces, sampling is steered towards the surface normal.

The path tracer utilizes Russian roulette for probabilistic path termination, selectively discarding paths with low expected radiance contribution. The samples are then averaged to compute the final pixel color.

\subsubsection{Render Pipeline}

The output of the path tracing compute shader is a texture, which is then passed to a rasterizer. The rasterizer renders the texture to the canvas using a fullscreen quad consisting of two triangles. Additionally, it applies tone mapping based on the Khronos PBR neutral tone mapper \cite{pbrNeutralToneMapping}.

After each sampling iteration, the current result is visualized, resulting in a progressive rendering approach.

\subsection{Performance}

Real-time performance is a key aspect of web-based applications. The renderer can visualize complex models in near real-time, requiring approximately 10 ms per sample on suitable hardware, taking approximately two seconds to render an image with 200 samples. The number of samples needed for a high-quality image depends on the scene complexity; effects such as reflections require more samples to converge.

As indicated by Table~\ref{tab:cpuPerformance}, CPU performance for BVH construction varies significantly based on the model's complexity. Table~\ref{tab:gpuPerformance} shows that GPU performance for path tracing is more stable across different model complexities and constitutes the most computationally expensive part of the process.

\section{Discussion}

The defined approach provides a viable solution to address challenges encountered in industrial applications. The ability to render complex assemblies in real-time without the need for pregenerated artifacts facilitates the use of CAD models with manifold assembly configurations and customer-specific materials. Product configurators using WebGPU can present product variety in high visual fidelity while enabling an interactive end-user experience with arbitrary viewing angles.

Additionally, operating costs can be reduced by eliminating the need to host a rendering pipeline or remote rendering service.

\subsection{Real-Time Rendering}

Path tracing is computationally expensive and requires multiple samples per pixel to achieve accurate results. Consequently, the sampling process is noticeable during interactions with the scene. One technique to improve perceived interaction quality is to overlay the rendering with a rasterization preview during interactions.

To reduce the number of samples required for higher quality renderings, techniques like neural radiance caching (NRC) \cite{muller2021real} or ReSTIR \cite{restir} could be employed in future work. Additionally, denoising algorithms applied as post-processing steps could enhance the quality of the results. Options include blockwise multi-order feature regression (BMFR) \cite{blockwise-multi-order-regresssion-for-rt-pt} and Open Image Denoise (OIDN) \cite{openImageDenoise}.

Further improvements can be achieved by aligning glTF PBR with OpenPBR, focusing on real-time rendering. This alignment could reduce the computational resources required for surface shading.

\subsection{Production Readiness of WebGPU}

WebGPU's production readiness remains limited due to incomplete support in Safari and Firefox. Both browsers have announced plans for implementation \cite{SafariWebGPUSupport,FirefoxWebGPUSupport}. Thanks to the extensive conformance test suite \cite{WebGPUConformanceTestSuite}, it is more likely that the different implementations will be compatible with each other.

The main browser supporting WebGPU to date is Chrome, which shipped WebGPU for general use on desktops in May 2023 \cite{ChromeWebGPUSupport}. Since January 2024, WebGPU has also been supported on modern Android devices \cite{ChromeAndroidWebGPUSupport}. This makes it straightforward to use WebGPU on most modern devices, with the notable exception of Apple iOS and iPadOS devices.

\subsection{Future Work}

Future work may include supporting additional OpenPBR features, implementing denoising techniques, integrating rendering techniques such as depth of field and volumetric path tracing, enhancing performance, and comparing the renderer with alternative web-based path tracers. The open-source nature of the project facilitates extension and serves as inspiration for other initiatives.

\begin{acks}
We thank the reviewers for their valuable feedback and suggestions. We also thank EAO and Intelliact for permitting the use of production CAD models, enabling the verification of a real-world use case.
\end{acks}

\newpage

\bibliographystyle{ACM-Reference-Format}
\bibliography{strahl-bibliography}

\newpage
\appendix
\section{Performance Measurements}

\begin{figure}[H]
  \includegraphics[width=0.9\columnwidth]{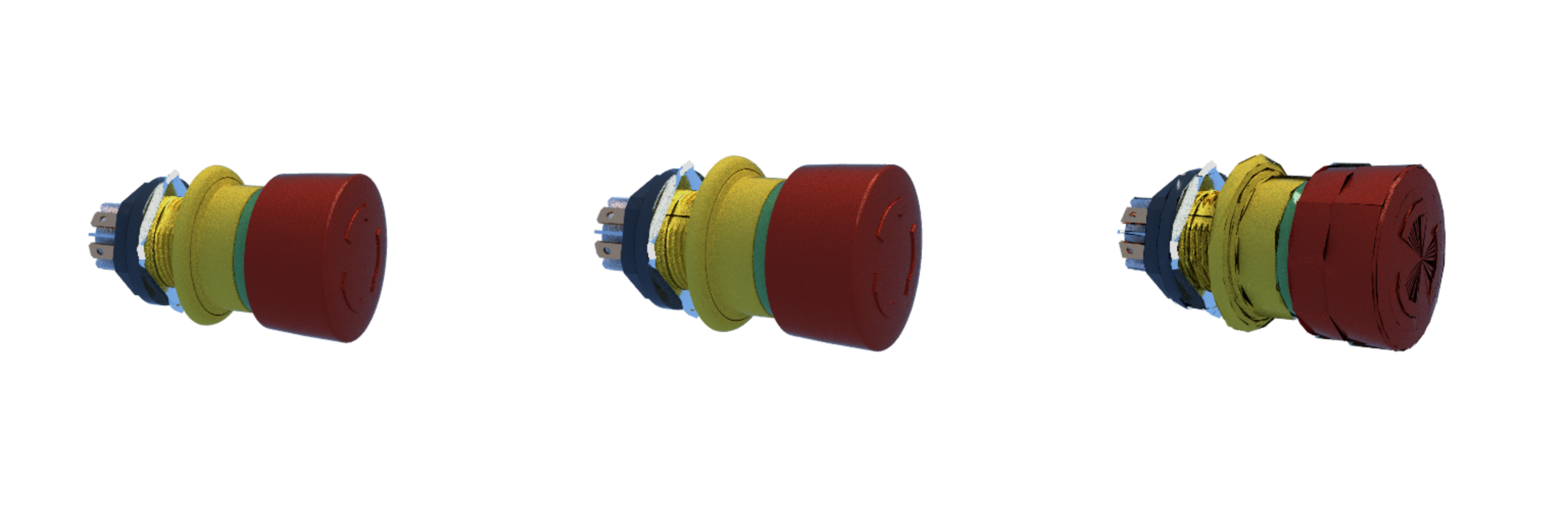}
  \caption{The three LOD artifacts, from left to right: 1,068,735 triangles, 106,873 triangles, 10,687 triangles. The left and middle figures share similar visual fidelity characteristics.}
  \Description{Three images showing a pushbutton model with different levels of detail from most detailed to least detailed. The difference between the most detailed and middle detail is minimal, while the least detailed version is significantly simplified.}
  \label{fig:benchmark-models}
\end{figure}

The measurements are split into two parts: CPU performance and compute pipeline GPU performance, as described in Figure~\ref{fig:path-tracer}. CPU performance excludes loading the glTF, which is heavily dependent on bandwidth and focuses on the BVH construction. For the GPU phase, a total of 100 samples per pixel with a ray depth of five was used. The image was rendered in Chrome 126 at a resolution of 512$\times$512 pixels. Experiments were conducted with different model complexities. The simplified versions are decimated meshes of the original, which consists of roughly one million triangles. The LOD artifacts are shown in Figure~\ref{fig:benchmark-models}. The first two levels are intended to be visually similar, while the third level is a simplified version intended to demonstrate the effect of non-manifold geometry for ray tracing.

The machine specifications used for the measurements are:

\begin{itemize}
  \item {Apple M1 Max}: MacBook Pro with Apple Silicon M1 Max 
  \item {AMD/NVIDIA}: AMD Ryzen 5 5600X and NVIDIA GeForce RTX 3080 (10 GB)
\end{itemize}

The results for CPU performance were recorded using the Performance API and are presented in Table~\ref{tab:cpuPerformance}. The results for GPU performance were recorded using WebGPU timestamp queries and are presented in Table~\ref{tab:gpuPerformance}. All measurements were calculated using the mean time of 30 benchmark samples, with a 95\% confidence interval given as $\pm$ standard deviation from the mean time in milliseconds.

\begin{table}[H]%
  \caption{BVH setup time based on model complexity}
  \label{tab:cpuPerformance}
  \begin{tabular}{rrr}
    \toprule
    Triangles   & Apple M1 Max    & AMD/NVIDIA \\ \midrule
    1,068,735     & 327.57 ms $\pm$ 0.56 ms     & 383.10 ms $\pm$ 4.06 ms \\
    106,873     & 45.49 ms $\pm$ 0.31 ms    & 41.77 ms $\pm$ 0.43 ms \\
    10,687     & 10.53 ms $\pm$ 0.24 ms    & 8.43 ms $\pm$ 0.22 ms \\
    \bottomrule
  \end{tabular}
\end{table}%

\begin{table}[H]%
  \caption{GPU path tracer time based on model complexity}
  \label{tab:gpuPerformance}
  \begin{tabular}{rrr}
    \toprule
    Triangles   & Apple M1 Max    & AMD/NVIDIA \\ \midrule
    1,068,735     & 2,319.45 ms $\pm$ 11.14 ms    & 1,058.73 ms $\pm$ 59.47 ms \\
    106,873     & 1,992.11 ms $\pm$ 8.49 ms    & 790.20 ms $\pm$ 3.79 ms\\
    10,687     & 2,031.38 ms $\pm$ 9.87 ms    & 790.83 ms $\pm$ 4.18 ms \\
    \bottomrule
  \end{tabular}
\end{table}%

The path tracer has not been optimized for performance. These measurements are intended to provide a baseline for future optimizations.

\end{document}